\documentclass{SPW9_proc}
\usepackage{bm}
\begin{document}
\title{Resonance Line Polarization in Spherically 
Symmetric Moving Media: a Parametric Study}
%
%\author[1,*]{A. Megha}
%\author[1]{M. Sampoorna}
%\author[1,2]{K. N. Nagendra}
%\author[3]{L. S. Anusha}
%\author[1,4,5]{K. Sankarasubramanian}
\author{A. Megha,$^{1,*}$ M. Sampoorna,$^{1}$ K. N. Nagendra,$^{1,2}$
L. S. Anusha,$^{3}$ and \\ K. Sankarasubramanian$^{1,4,5}$}
%\affil[1]{Indian Institute of Astrophysics, Koramangala, Bengaluru, India}
%\affil[2]{Istituto Ricerche Solari Locarno, Locarno Monti, Switzerland}
%\affil[3]{Max-Planck-Institut f\"ur Sonnensystemforschung, 
%Justus-von-Liebig-Weg 3, D-37077, G\"ottingen, Germany}
%G\"ottingen, Germany}
%\affil[4]{Space Astronomy Group, U. R. Rao Satellite Centre, ISITE Campus, 
%Bengaluru}
%\affil[5]{CESSI, IISER, Kolkata, India}
%\affil[*]{\textit{Email:} megha@iiap.res.in}
%***************************************************************
 \affil{$^1$Indian Institute of Astrophysics, Koramangala, Bengaluru, India}
 \affil{$^2$Istituto Ricerche Solari Locarno, Locarno Monti, Switzerland}
 \affil{$^3$Max-Planck-Institut f\"ur Sonnensystemforschung, 
 G\"ottingen, Germany}
 \affil{$^4$Space Astronomy Group, U. R. Rao Satellite Centre, ISITE Campus, 
 Bengaluru}
 \affil{$^5$CESSI, IISER, Kolkata, India}
 \affil{$^{*}$\textit{Email:} megha@iiap.res.in}
\runningtitle{Resonance line polarization in spherically
symmetric moving media}
\runningauthor{A.~Megha et al.}
\firstpage{1}
\maketitle
\begin{abstract}
In the present paper we consider the problem of resonance line polarization
formed in the spherically symmetric expanding atmospheres. 
For the solution of the concerned polarized
transfer equation we use the comoving frame formulation, and apply the 
Accelerated Lambda Iteration (ALI) method. 
We restrict ourselves to the non-relativistic regime of velocities
wherein mainly Doppler shift effects are significant. 
For our studies, we consider
the scattering on a two-level atom, including the effects 
of partial frequency redistribution (PFR). 
We present the dependence of linearly polarized profiles on different 
atmospheric and atomic parameters.
\end{abstract}
\section{Introduction}
The spectroscopic observations of different classes of astrophysical
objects like supergiants, Wolf-Rayet stars, novae, and supernovae indicate the
existence of high-velocity outward gas flows. 
The spectral line profiles of these objects are of P Cygni 
type \citep[]{beals1931} 
with redshifted emission and blueshifted absorption indicating the
rapid outflow of matter in the outer layers of their atmospheres.
These outward gas flows lead to extended atmospheres. 
Even small extensions present in solar type stars are highly structured 
which give rise to significant observable effects. 
Therefore a precise treatment of such problems require going beyond the 
plane-parallel model.

The detailed discussions on techniques for solving the problem of line
formation in extended and expanding atmospheres in both observer's frame
and comoving frame (CMF) are described in 
\citet[and references therein]{hm2014}.
These earlier works considered only the intensity profiles 
accounting for either
complete frequency redistribution (CFR) or PFR in scattering. 
In \cite[]{megha2019}, we extended this problem to include the resonance line 
polarization. To this end we
extended CMF-ALI method of \cite[]{hb2004} to include polarization and PFR. 
For our studies we considered the CMF transfer equation in the non-relativistic
limit (i.e., the advection and aberration terms were excluded and only 
Doppler shift effects were retained). Here we use the CMF-PALI 
method presented in \cite[]{megha2019} to understand the quantitative 
behavior of the linear polarization when the basic model parameters are 
varied systematically one at a time keeping the other parameters as constants. 
%%%%%%%%%%%%%%%%%%%%%%%%%%%%%%%%%%%%%%%%%%%%%%%%%%%%%%%%%%%%%%%
 \begin{figure*}%[!ht]
         \begin{center}
\includegraphics[width=10cm,height=6.cm]{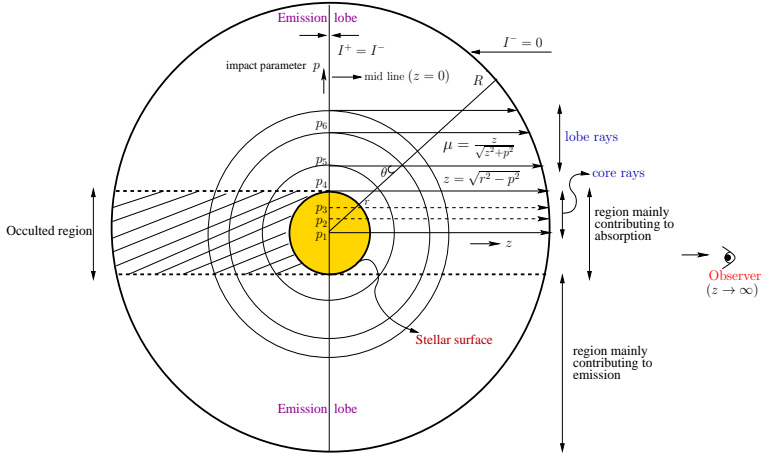}
\caption{The ($p,z$) representation used to solve the spherically 
symmetric radiative transfer equation.}
 \label{geom}
         \end{center}
 \end{figure*}
%%%%%%%%%%%%%%%%%%%%%%%%%%%%%%%%%%
\section{The comoving frame polarized transfer equation}
In order to effectively treat the outward peaking of the radiation field
in a spherically symmetric atmosphere,
we consider the transfer equation in $(p,z)$ representation
\citep[]{hr1971}, where $p$ is the impact parameter of the
ray and $z$ is the distance measured along it (see Figure \ref{geom}). 
In this representation, the polarized CMF transfer equation,
under the non-relativistic limit is given by 
\begin{equation}
\pm \frac{\partial \boldsymbol{\mathcal{I}}^{\pm}(z,p,x)}{\partial \tau(z,x)} = 
\boldsymbol{\mathcal{I}}^{\pm}(z,p,x)-\boldsymbol{\mathcal{S}}(z,x)
%- {\boldsymbol{\tilde{\mathcal{S}}}(z,p,x)},
+ \frac{a(r,p)}{\chi(r,x)}
\frac{\partial \boldsymbol{\mathcal{I}}
 ^{\pm}(z,p,x)}{\partial x},
\label{comoving-rte}
\end{equation}
where $+$ and $-$ stand for outgoing and incoming rays respectively, 
$\boldsymbol{\mathcal{I}^{\pm}}=[I_0^{0,\pm},I_0^{2,\pm}]^{\rm{T}}$ is 
a 2-component irreducible specific
intensity vector \citep[]{f2007}, and $x=(\nu-\nu_0)/\Delta\nu_{\rm D}$ (with 
$\nu_0$ being the line center frequency and $\Delta\nu_{\rm D}$ the Doppler 
width). The monochromatic optical depth along the tangent ray is defined as 
$d\tau=[\varphi(x)+\beta_c]d\tau_r/\mu$ (with $r$ being the radial distance),
 where $d\tau_r=-\chi_l(r)dr$ is the 
radial optical depth with $\chi_l(r)$ being the line averaged 
absorption coefficient, $\mu(r,p)=\sqrt{1-(p^2/r^2)}$, and 
$\beta_c=\chi_c(r)/\chi_l(r)$ with $\chi_c(r)$ 
being the continuum absorption coefficient. The total absorption coefficient
$\chi(r,x)= {\chi_l(r)\varphi(x)+\chi_c(r)}$, where the 
line absorption 
profile function $\varphi(x)$=$H(a,x)$ is a Voigt function, with $a$ the 
damping parameter. In the CMF, $\varphi(x)$ is angle-independent. 
All the effects of the velocity field are contained in the last term of 
Equation (\ref{comoving-rte}), wherein
\begin{equation}
{a(r,p)} = (1-\mu^2) \frac{V}{r}+\mu^2
\frac{dV}{dr}.
\label{ax}
\end{equation}
Here $V=v_r/v_{\rm th}$ is the non-dimensional velocity, with $v_r$ the 
radial velocity and $v_{\rm th}$ the thermal velocity.
The CMF total source vector is given by 
\begin{equation}
{{\boldsymbol{\mathcal S}}(z,x)}=
{\varphi(x){\boldsymbol{\mathcal S}}_{l}(z,x)+
\beta_c {\boldsymbol{\mathcal S}}_{c} \over
\varphi(x)+\beta_c},
\end{equation}
where the unpolarized continuum source vector ${\boldsymbol{\mathcal S}}_{c}=
B_{\nu_0}{\boldsymbol{\mathcal U}}$, with $B_{\nu_0}$ the Planck function at 
the line center, and ${\boldsymbol{\mathcal U}}=(1,0)^{\rm T}$. For a two-level 
atom with infinitely sharp and unpolarized lower level, the line source vector
has the form
\begin{equation}
{\boldsymbol{\mathcal S}}_{l}(z,x) = 
\epsilon B_{\nu_0}{\boldsymbol{\mathcal U}} +
 \int _{-\infty}^{+\infty} dx^\prime{\frac{1}{2}}\int_{-1}^{+1} 
d\mu^\prime
{\frac{{\boldsymbol{\mathcal R}} (x,x^\prime)}{\varphi(x)}}\,
{\bf \Psi}(\mu^\prime){\boldsymbol{\mathcal I}}(\tau, \mu^\prime, x^\prime). 
\end{equation}
Here $\epsilon$ is the thermalization parameter, ${\boldsymbol{\mathcal R}}$ 
the $2\times 2$ non-magnetic angle-averaged PFR matrix 
\citep[]{b1997}, and ${\bf \Psi}$ the Rayleigh phase matrix in 
the irreducible basis \citep[]{f2007}. 

We solve Equation (\ref{comoving-rte}) by applying the polarized ALI method
(see, e.g., the reviews by \citealt[]{knn2003, knn2014}). The details of this 
method are discussed in \cite[]{megha2019}. Hence we do not repeat them here.
%%%%%%%%%%%%%%%%%%%%%%%%%%%%%%%%%%%%%%%%%%%%%%%%%%%%%%%%%%%%%%%%%%%
\section{Model parameterization}
For our studies we consider the following set of `standard model parameters'. 
An isothermal, spherically symmetric atmosphere 
with inverse square law opacity distribution (i.e., 
$\chi_{l,c}\propto r^{-n}$, here 
the power law opacity index $n$=2), a frequency averaged total radial line
optical thickness of 
$T$, and an outer radius $R$ is considered. 
We use reflecting boundary condition, namely 
$\boldsymbol{\mathcal{I}}^{+}(\tau=T,p,x)=
\boldsymbol{\mathcal{I}}^{-}(\tau=T,p,x)$ at the lower boundary and  
$\boldsymbol{\mathcal{I}}^{-}(\tau=0,p,x)=0$ at the outer boundary. 
For discretization of radius $r$, the impact parameter $p$, angle $\mu$, 
frequency $x$, and depth grid $\tau_r$ we have followed 
\cite[]{als2009}. The `standard model parameters' are:
$T=10^6$, $R=30$, $\beta_c=10^{-6}$,
$\epsilon=10^{-4}$, $a=10^{-3}$, $B_{\nu_0}=1$, and no elastic collisions.
%We use these values of the parameters as the standard values. 
For our studies we consider both static ($V(r)$=0) and a 
constantly moving atmosphere with $V(r)$=3. We show all the results for a
fixed line-of-sight of $\mu$=0.11.
The effects of extendedness $R$ 
and the elastic collisions 
on the linearly polarized line profiles formed in 
static media and in the presence of velocity fields are presented in
\cite[]{megha2019}. Following 
\citet[]{knn1988,knn1994,knn1995} here we present the dependence
of linearly polarized
profiles on $a$, $\epsilon$, $\beta_c$, $T$, 
and the power law opacity index $n$, when they are varied one at a time
around the standard model parameters, while
keeping the other parameters as constants.
%%%%%%%%%%%%%%%%%%%%%%%%%%%%%%%%%%%%%%%%%%%%%%%%%%%%%%%%%%%%%
\section{Results and discussions}
%%%%%%%%%%%%%%%%%%%%%%%%%%%%%%%%%%%%%%%%%%%%%%%%%%%%%%
 \begin{figure*}%[!ht]
         \begin{center}
\includegraphics[width=5cm,height=8.5cm]
{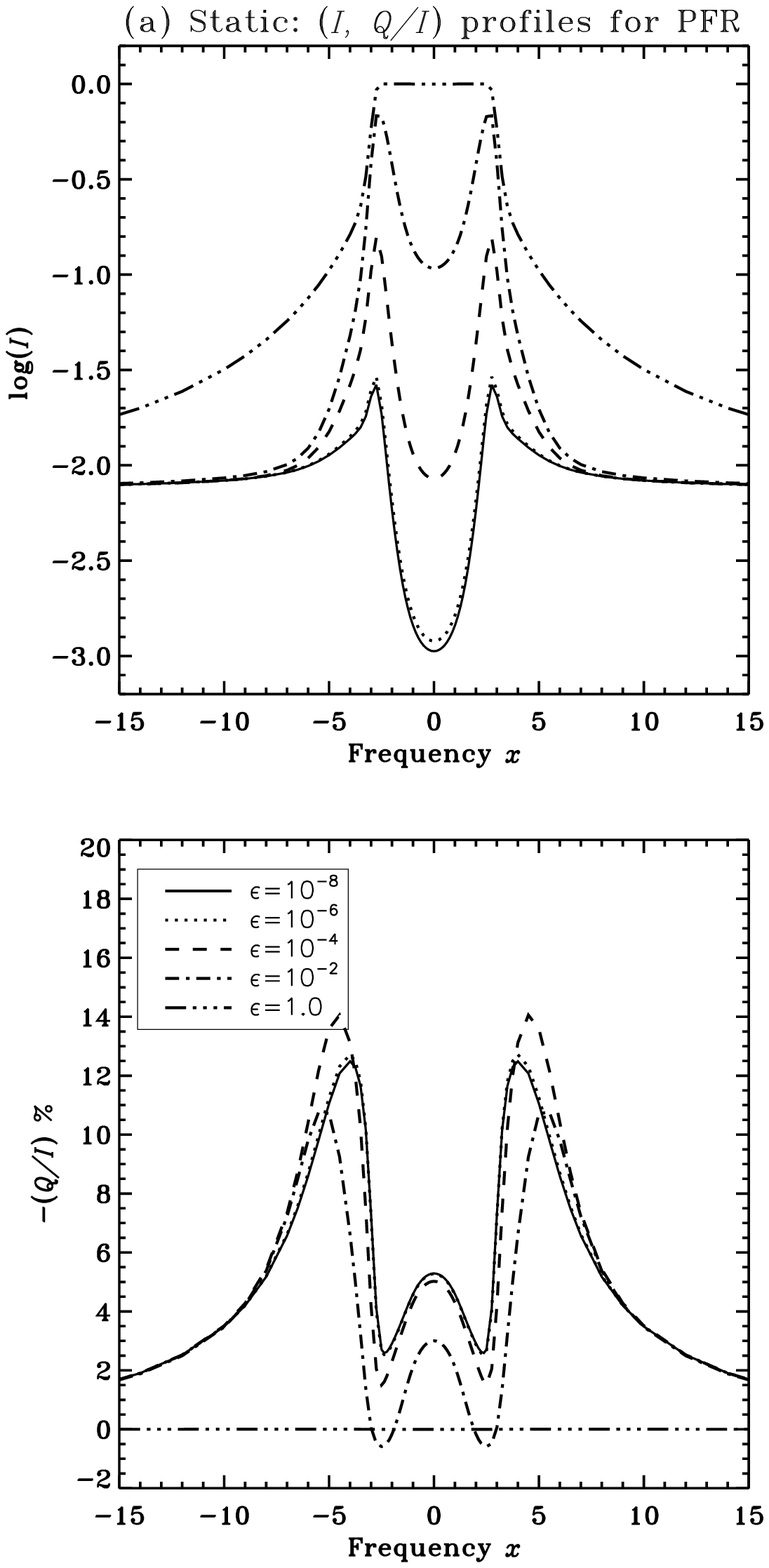}
\includegraphics[width=5cm,height=8.5cm]
{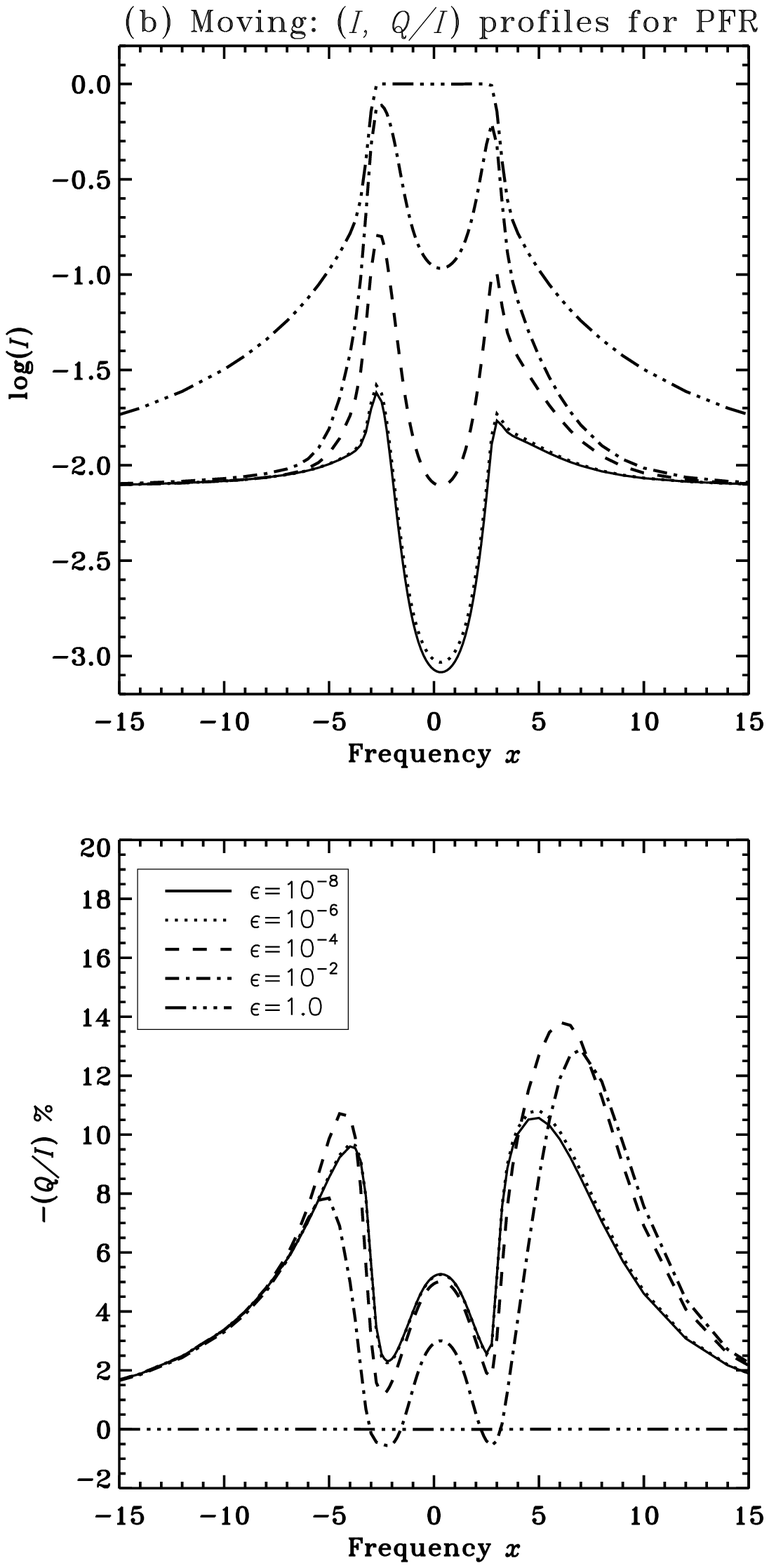}
 \caption{Emergent PFR $(I, Q/I)$ profiles from a spherically symmetric
static (panel a) and a constantly moving atmosphere
with $V(r)$=3 (panel b) for varying values of thermalization parameter
$\epsilon$ (which are given in the inset box).}
 \label{epsilon}
         \end{center}
 \end{figure*}
%%%%%%%%%%%%%%%%%%%%%%%%%%%%%%%%%%%%%%%%%%%%%%%%%%%%%%%%%%%
\subsection{Dependence on thermalization parameter $\epsilon$}
Figure \ref{epsilon} shows the emergent PFR $(I, Q/I)$ profiles 
from a spherically symmetric static (panel a) and a constantly moving 
atmosphere with $V(r)$=3 (panel b) for varying values of 
thermalization parameter $\epsilon$ in the range [$10^{-8}$, 1]. 
$\epsilon$ gives the probability that a photon is destroyed 
during scattering process due to collisional de-excitation. 
Here $\epsilon$=1
refers to pure LTE case and other values refer to non-LTE case. As the 
value of $\epsilon$ increases from $10^{-8}$ to 1, the thermal
coupling of photons with the continuum also increases due to increase in the
number of collisions in the medium. 
For $\epsilon=10^{-8}$, the intensity shows a self-absorbed 
symmetric emission profile
for the static case (see Figure \ref{epsilon}a). As the $\epsilon$ value 
increases from $10^{-8}$ to 1, the intensity around the line core region 
increases. Also self-absorbed part of the profile becomes shallower
and finally disappears for $\epsilon=1$. 
Furthermore for $\epsilon=1$, the symmetric
emission peaks around $x=\pm$2.25 disappear as there is no
scattering contribution from the extended lobes. 
The polarization profile (Figure \ref{epsilon}a, lower panel) 
for $\epsilon=10^{-8}$ shows a triple peak
structure due to PFR. As $\epsilon$ value increases, the magnitude of 
polarization in the line core region decreases 
due to a decrease in the number of scatterings and the profile 
becomes completely flat with zero polarization for $\epsilon=1$. 
However at the PFR wing peaks, we see that $Q/I$ is nearly the same for
$\epsilon$ between $10^{-8}$ and $10^{-6}$, then increases for 
$10^{-5}\leq \epsilon \leq 10^{-4}$, and then decreases for 
$\epsilon \geq 10^{-3}$. To understand this we recall that in a spherical
atmosphere with $B_{\nu_0}=1$ photons are created substantially closer to the
surface (see \citealt[]{kh1974}), from where they escape much more readily.
With increase in $\epsilon$ the photon creation rate increases while mean
number of scatterings decrease. For $\epsilon \leq 10^{-6}$, mean number
of scatterings are nearly the same (see Table IV of 
\citealt[]{kh1974} which corresponds to the case of $n=0$) so that
$I$ and $Q/I$ are nearly the same for $\epsilon$ between $10^{-8}$ 
and $10^{-6}$. 
For $10^{-5}\leq \epsilon \leq 10^{-4}$, although the mean number of 
scatterings decrease relatively, since the photons are created near the
surface a moderate number of scatterings would result in larger values of 
$Q/I$. This is particularly the case for highly extended atmosphere
($R=30$) considered in this paper. For $\epsilon > 10^{-4}$ mean number
of scatterings decrease considerably resulting in decreasing values of 
$Q/I$ as in planar atmospheres. Our numerical studies show that, such a
dependence of $Q/I$ on $\epsilon$ is seen for $R > 10$, while for $R \leq 10$
the $Q/I$ monotonically decreases with increasing $\epsilon$ not only 
in the line core but also in the PFR wings. 

In the presence of velocity fields the line profiles 
are asymmetric about the line center (see Figure \ref{epsilon}b). 
In particular
the red and blue wing PFR peaks in $Q/I$ get highly affected
in the presence of velocity gradients. We recall that although we have 
considered a constant velocity field, the Doppler shift $\mu(r,p)V(r)$ along
a given impact parameter ray changes (as $\mu$ varies due to sphericity 
effects) thereby producing a velocity gradient in the $z$-direction.
The dependence of $I$ and $Q/I$ profiles on $\epsilon$ for non-zero velocity
field is similar to the corresponding static case. Apart from producing
Doppler shift, the velocity fields modify the source function gradient
thereby either enhancing or reducing the anisotropy of the radiation field.
In a spherical atmosphere this modification is different for red and blue wing
PFR peaks of the $Q/I$ profile (due to sphericity effects and as the 
intensity is a self-absorbed profile), which results in the high
asymmetry noted above. A more detailed discussion can be found in 
\cite[]{megha2019}.
%%%%%%%%%%%%%%%%%%%%%%%%%%%%%%%%%%%%%%%%%%%%%%%%%%%%%%%%%%%%%%%%%%%%%%
\begin{figure*}
         \begin{center}
\includegraphics[width=5cm,height=8.5cm]
{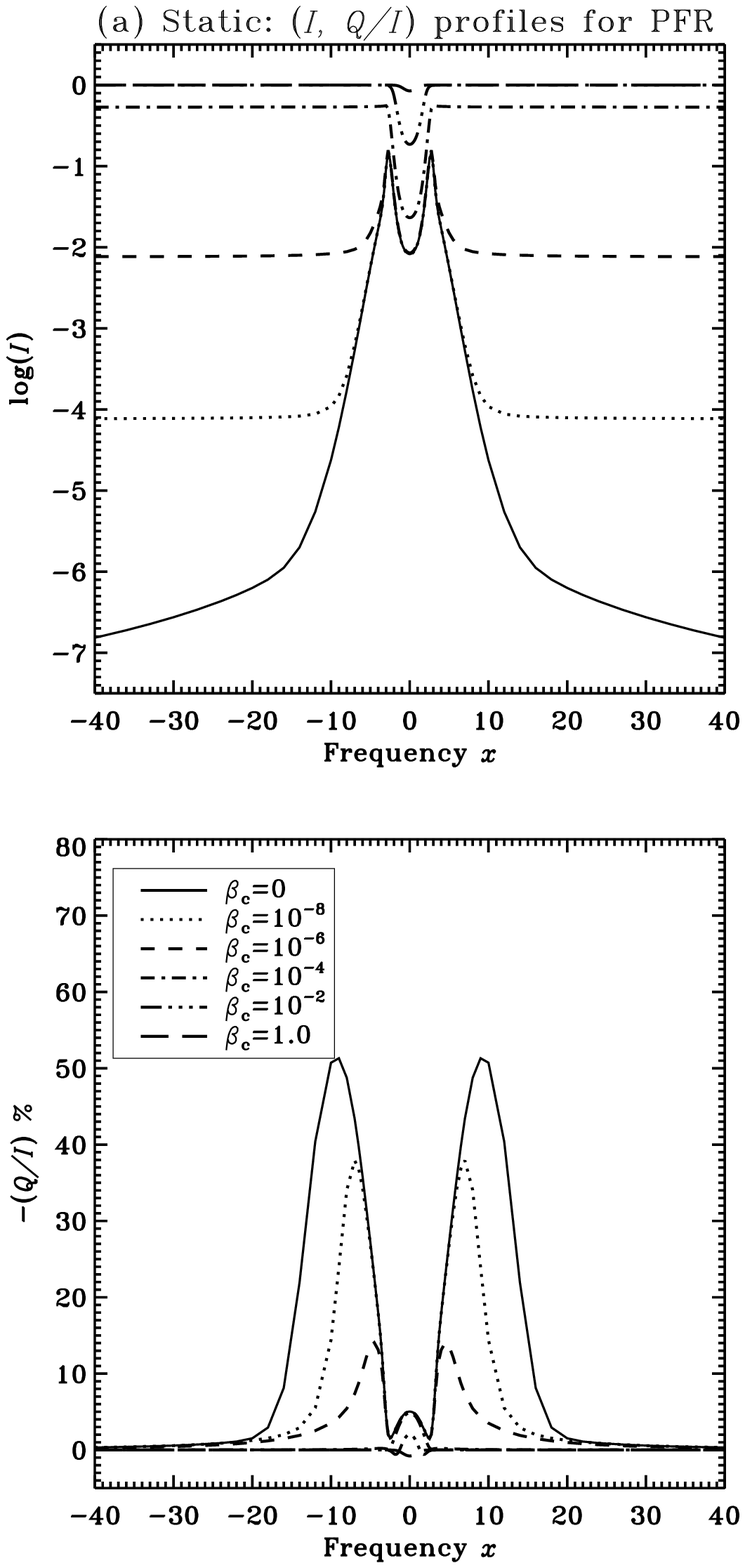}
\includegraphics[width=5cm,height=8.5cm]
{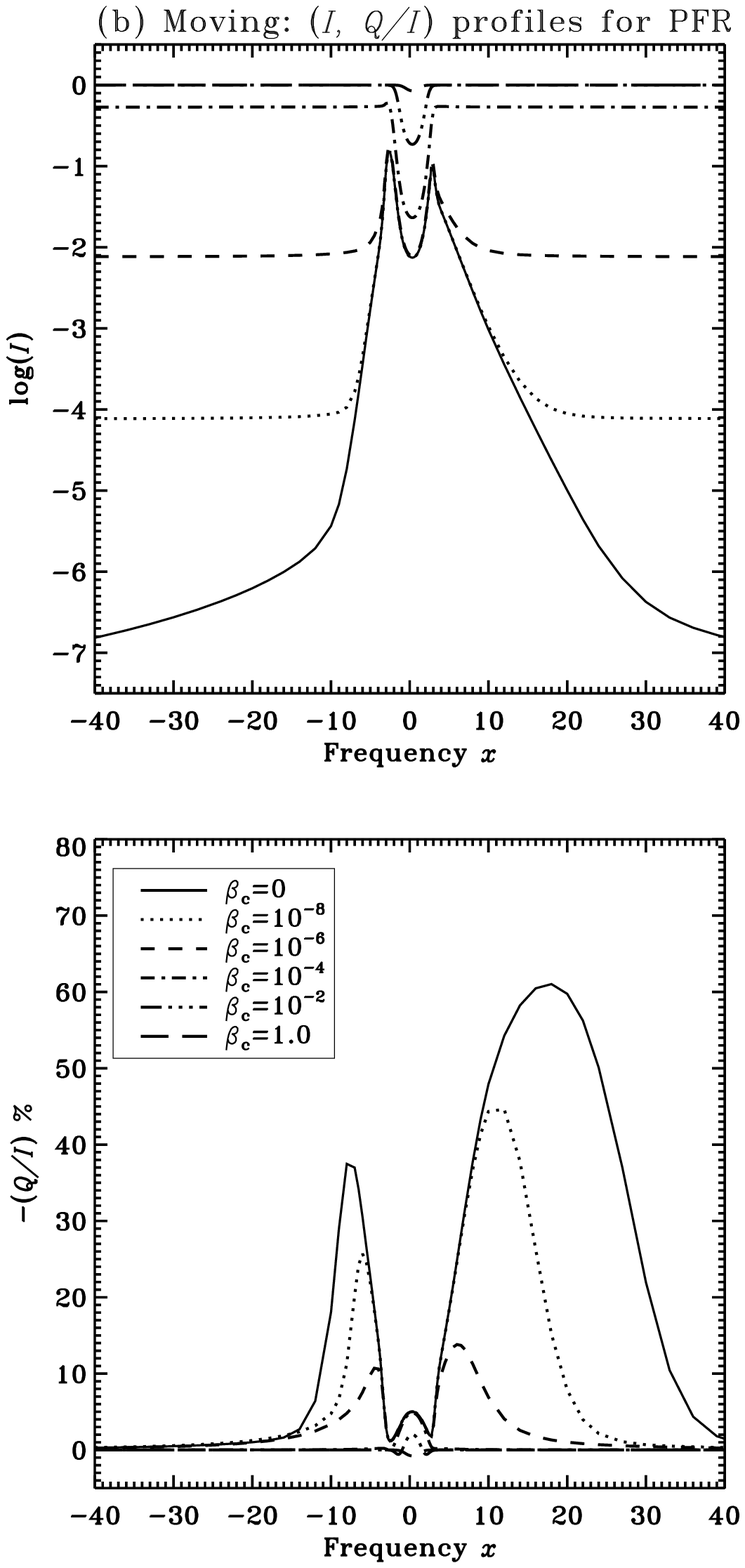}
 \caption{Emergent PFR $(I, Q/I)$ profiles from a spherically symmetric
static (panel a) and a constantly moving atmosphere
with $V(r)$=3 (panel b) for varying values of continuous absorption co-efficient
$\beta_c$ (which are given in the inset box).}
 \label{betac}
         \end{center}
 \end{figure*}
%%%%%%%%%%%%%%%%%%%%%%%%%%%%%%%%%%%%%%%%%%%%%%%%%%%%%%%%%%%%%
\subsection{Dependence on continuous absorption parameter $\beta_c$}
Figure \ref{betac} shows the emergent PFR $(I, Q/I)$ profiles 
from a spherically symmetric static (panel a) and a constantly moving 
atmosphere with $V(r)$=3 (panel b) for varying values of continuous 
absorption parameter $\beta_c$ in the range [0, 1]. 
$\beta_c$ is the ratio of continuum opacity to frequency averaged
line opacity both of which are taken to follow the inverse square law.
Hence $\beta_c$ remains constant throughout the spherical atmosphere. 
Here the continuum optical depth which is $\beta_c T$ is different for 
different models.
$\beta_c=0$ corresponds to a pure line case. Here 
the intensity shows self absorbed emission profile, with the wings falling
sharply towards zero due to absence of absorption. 
With the increase of $\beta_c$ from 0 to $10^{-6}$ the intensity
does not change much in the line core region but the increase in the 
value of $\beta_c$ contributes to the continuum leading to larger values
of intensity in the wings. Consequently the emission
peaks slowly decrease in their height relative to the continuum 
and finally cease to exist for $\beta_c=10^{-4}$, for which the intensity
now exhibits an absorption profile.
For $\beta_c=0$, the polarization shows triple peaks due to PFR with highly
enhanced wing peaks. 
The $Q/I$ at the wing PFR peaks is of the order of 50 $\%$.
This is because of the strong outward peaking of the radiation field in a 
spherically symmetric atmosphere, which makes the radiation field highly 
anisotropic. With the increase in the value of $\beta_c$,
the magnitude of polarization reduces as the contribution from the
unpolarized continuum photons
dilute the polarized radiation field. Especially the wing PFR peaks sharply 
fall down and they disappear for $\beta_c=10^{-4}$ and 
with a sign reversal for $\beta_c=10^{-2}$.
This is because the unpolarized continuum source function becomes equal to
or larger than the line source function at progressively smaller 
frequencies (\citealt[see also \citealt{knn1995}]{faurobert1988}).
For $\beta_c=1$, the polarization nearly becomes zero with weak negative
polarization in the line core. In the presence
of velocity field, the intensity for $\beta_c < 10^{-4}$ is strongly 
affected in the blue region. The effect of velocity field exists only
in the line core region of intensity profile with further increase in $\beta_c$
as the continuum absorption dominates over that of line.
The presence of velocity field has similar effects on the polarization
profiles with strong enhancement in the magnitude of blue wing PFR peak
for $\beta_c < 10^{-4}$. 
%%%%%%%%%%%%%%%%%%%%%%%%%%%%%%%%%%%%%%%%%%%%%%%%%%%%%%%%%%%%%%%%%%%%%%
\begin{figure*}
         \begin{center}
\includegraphics[width=5cm,height=8.5cm]
{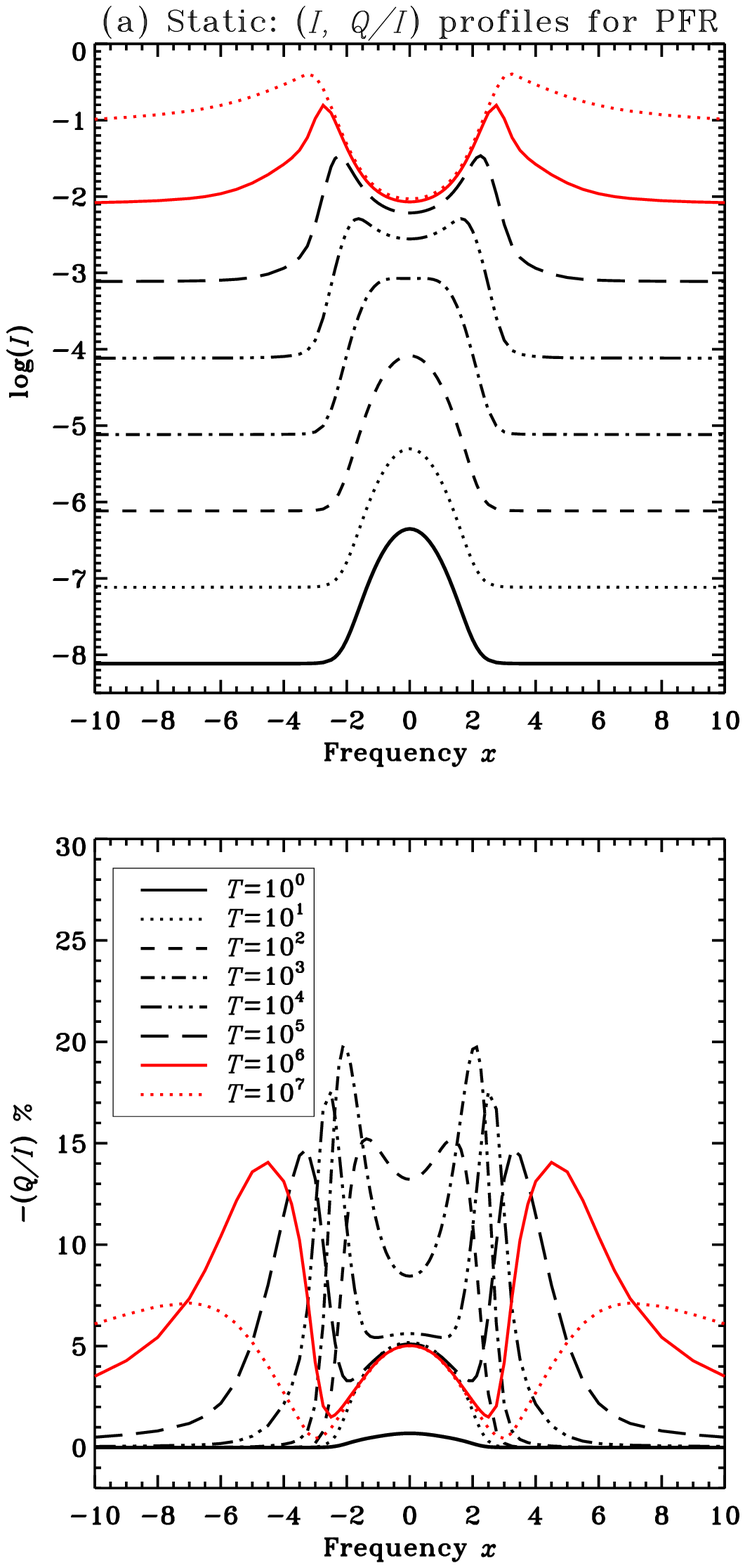}
\includegraphics[width=5cm,height=8.5cm]
{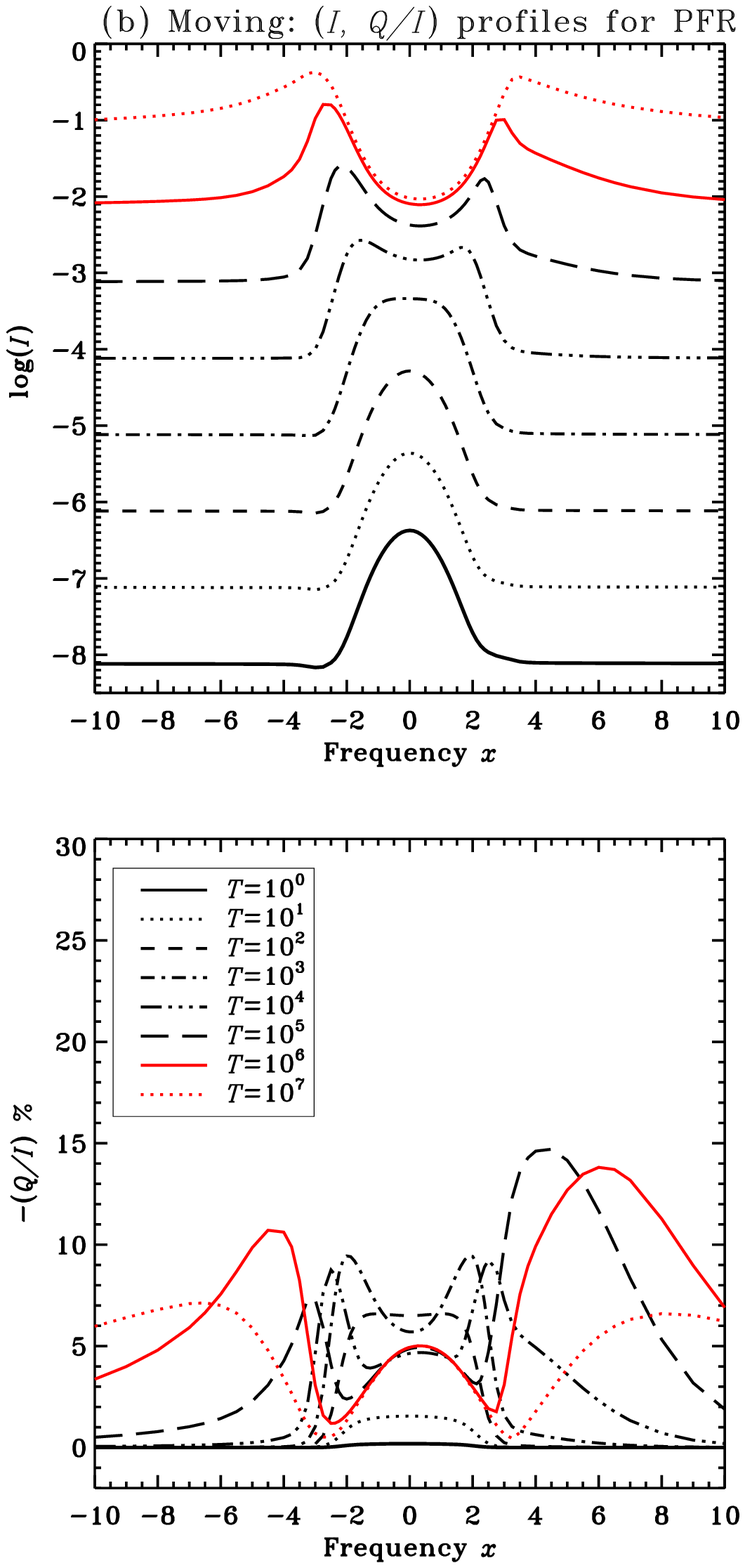}
 \caption{Emergent PFR $(I, Q/I)$ profiles from a spherically symmetric
static (panel a) and a constantly moving atmosphere
with $V(r)$=3 (panel b) for varying values of line averaged radial
optical thickness $T$ (which are given in the inset box).}
 \label{T}
         \end{center}
 \end{figure*}
%%%%%%%%%%%%%%%%%%%%%%%%%%%%%%%%%%%%%%%%%%%%%%%%%%%%%%%%%%%%%%%%%%%%%%
\subsection{Dependence on line averaged radial optical thickness $T$}
Figure \ref{T} shows the emergent PFR $(I, Q/I)$ profiles 
from a spherically symmetric static (panel a) and a constantly moving 
atmosphere with $V(r)$=3 (panel b) for varying values of 
optical thickness $T$ from 1 to $10^7$. Such a wide variation in $T$
allows us to sample effectively thin ($\epsilon T = 10^{-4}$) to thick 
($\epsilon T = 10^{3}$) atmospheres. The intensity is in emission
up to $T=10^3$ with increasing magnitude and line width. 
This is clearly an NLTE effect, as number of scatterings increase as $T$
increases. For $T > 10^4$, the medium starts to become effectively thick
particularly in the stellar core, thereby giving rise to self absorption
in the line core. Thus for $T$ in the range $10^{4}$ to 
$10^{7}$, the intensity shows self absorbed emission profile with increased
broadening. The linear polarization $Q/I$ profiles show a clear sensitivity
to variations in $T$. For $T$ from 1 to
10, the polarization is confined to the line core. 
This is because the medium is effectively thin ($\epsilon T \ll 1$).
With further increase in $T$ from $10^2$ to $10^5$, PFR wing
peaks start to appear with wing peaks shifting towards higher frequency
points. Furthermore for $T > 10^3$ the wing PFR peaks exhibit a progressive
decrease in magnitude.
This outward shift and decrease in magnitude are due to the larger optical
depths in the outer layers as the medium becomes more and more effectively 
thick with increasing $T$. For the same reason for 
$T \geq 10^5$, the central peak is also formed in $Q/I$. The 
velocity fields give rise to a 
Doppler blue shift and also an asymmetry about the line center
in the $(I,Q/I)$ profiles. The dependence of $I$ profiles on $T$ for 
non-zero velocity field is similar to the corresponding static case.
For $T \leq 10^4$, the $Q/I$ profiles with velocity fields exhibit a strong
Doppler dimming throughout the profile. This is because the corresponding 
intensity profiles are in emission except for $T=10^4$. For the same reason 
the $(I,Q/I)$ profiles are weakly asymmetric about the line center. 
For $T=10^4$
a shallow self-absorption in the line core region is seen in $I$ profile.
Consequently a slight asymmetry between the red and blue wing PFR peaks in
$Q/I$ is seen accompanied with slightly broadened blue wing PFR peak.
This asymmetry and broadening of blue wing PFR peak in $Q/I$ are enhanced
for $T > 10^4$, as for these cases intensity exhibits a more stronger 
self-absorption in the line core.
%%%%%%%%%%%%%%%%%%%%%%%%%%%%%%%%%%%%%%%%%%%%%%%%%%%%%%%%%%%%%%%%%%%%%%
%%%%%%%%%%%%%%%%%%%%%%%%%%%%%%%%%%%%%%%%%%%%%%%%%%%%%%%%%%%%%%%%%%%%%%
\begin{figure}
         \begin{center}
\includegraphics[width=5cm,height=8.5cm]
{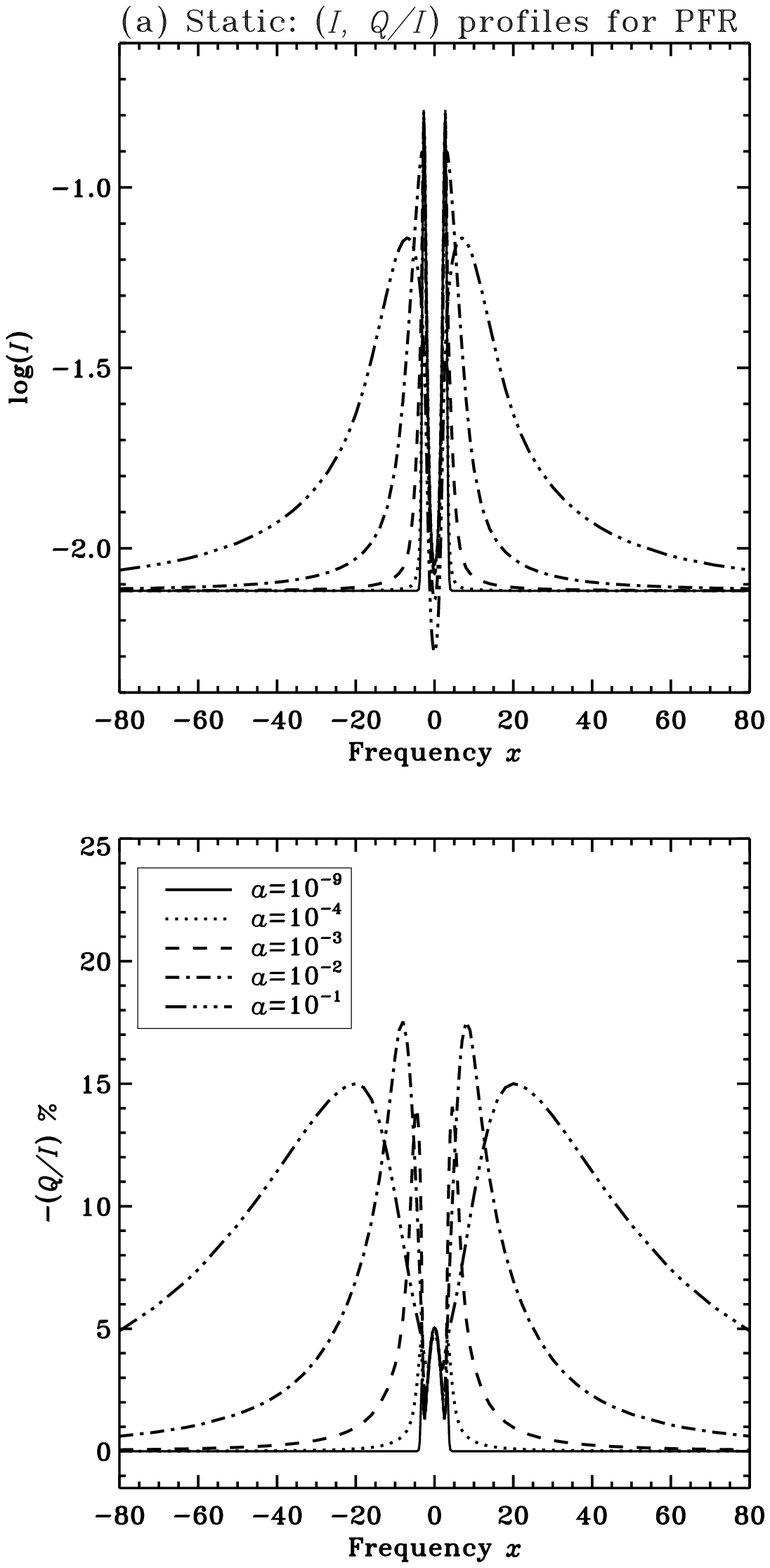}
\includegraphics[width=5cm,height=8.5cm]
{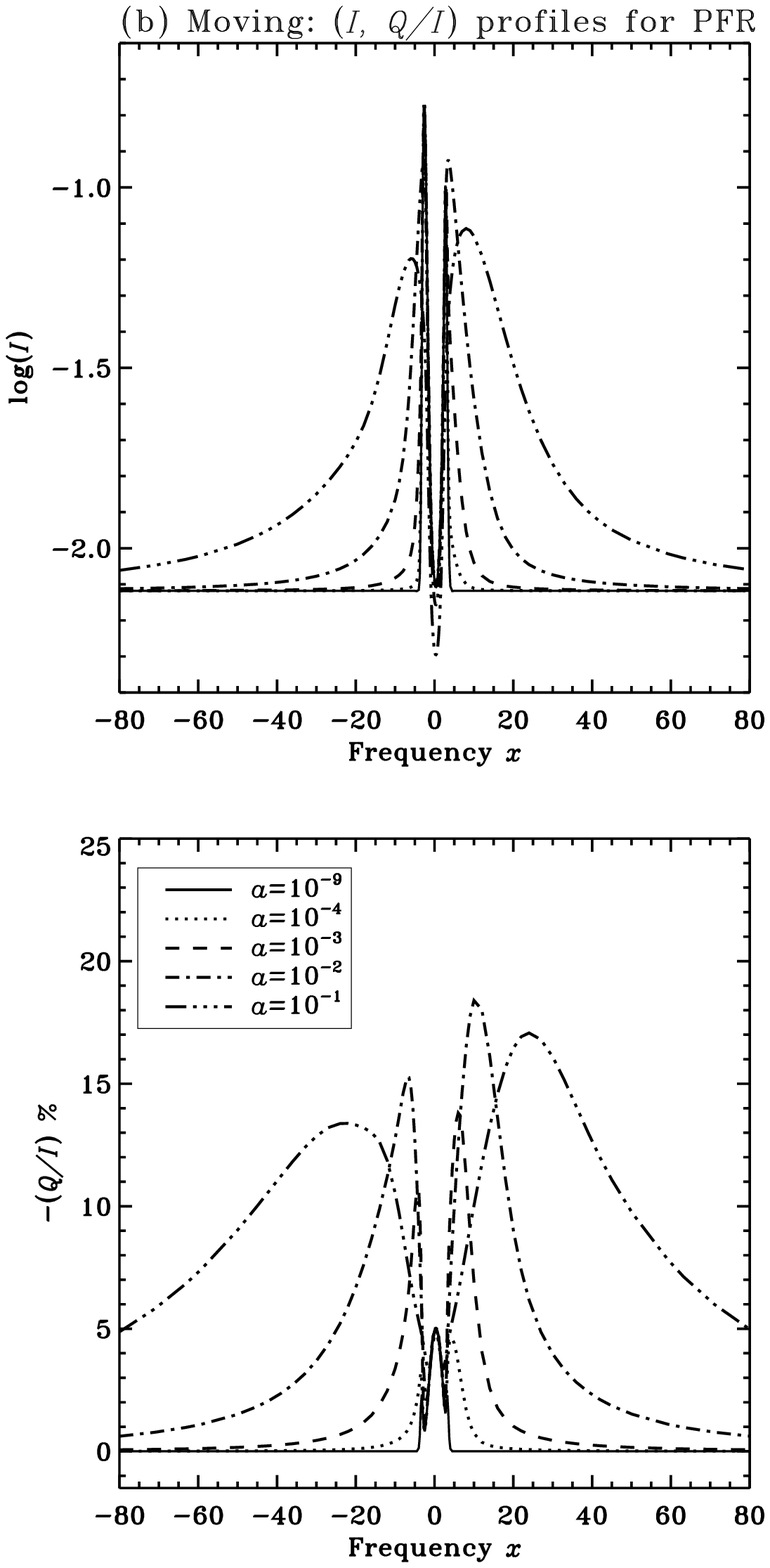}
 \caption{Emergent PFR $(I, Q/I)$ profiles from a spherically symmetric
static (panel a) and a constantly moving atmosphere
with $V(r)$=3 (panel b) for varying values of damping parameter
$a$ (which are given in the inset box).}
 \label{damping-param}
         \end{center}
 \end{figure}
%%%%%%%%%%%%%%%%%%%%%%%%%%%%%%%%%%%%%%%%%%%%%%%%%%%%%%%%%%%%%%%%%%%%%%
%%%%%%%%%%%%%%%%%%%%%%%%%%%%%%%%%%%%%%%%%%%%%%%%%%%%%%%%%%%%%%%%%%%%%%
\subsection{Dependence on damping parameter $a$}
Figure \ref{damping-param} shows the emergent PFR $(I, Q/I)$ profiles 
from a spherically symmetric static (panel a) and a constantly moving 
atmosphere with $V(r)$=3 (panel b) for varying values of 
damping parameter $a$ in the range $10^{-9}$ to $10^{-1}$. 
Since we neglect the effects of elastic collisions (namely,
$a=\Gamma_R/4\pi \Delta \nu_D$), the variation of $a$
is equivalent to the variation in radiative damping width $\Gamma_R$.
As the value of $a$ increases intensity profiles become more broader leading
to a strong damping wings. The linear polarization profiles
also exhibit a broadening with increasing $a$. In particular the wing PFR
peaks are much broader and also shift to larger frequency away from the 
line center. Moreover, amplitude of wing PFR peaks initially increase with 
increasing values of $a$ and then decreases for $a=0.1$ which is perhaps
due to excessive line broadening. We recall that 
with increasing values of $a$ the PFR becomes more and 
more important resulting in ($I,Q/I$) profiles shown in Figure 
\ref{damping-param}. 
This dependence of ($I,Q/I$) profiles on damping parameter $a$
is similar to the planar case (see e.g., Figure 8 of \citealt[]{ms2010}).
In the presence of
velocity field the asymmetries are present in $(I, Q/I)$ profiles, while the 
dependence on variation of $a$ is similar to the corresponding static case.  
%%%%%%%%%%%%%%%%%%%%%%%%%%%%%%%%%%%%%%%%%%%%%%%%%%%%%%%%%%%%%%%%%%%%%%
\begin{figure*}
	 \begin{center}
\includegraphics[width=5cm,height=8.5cm]
{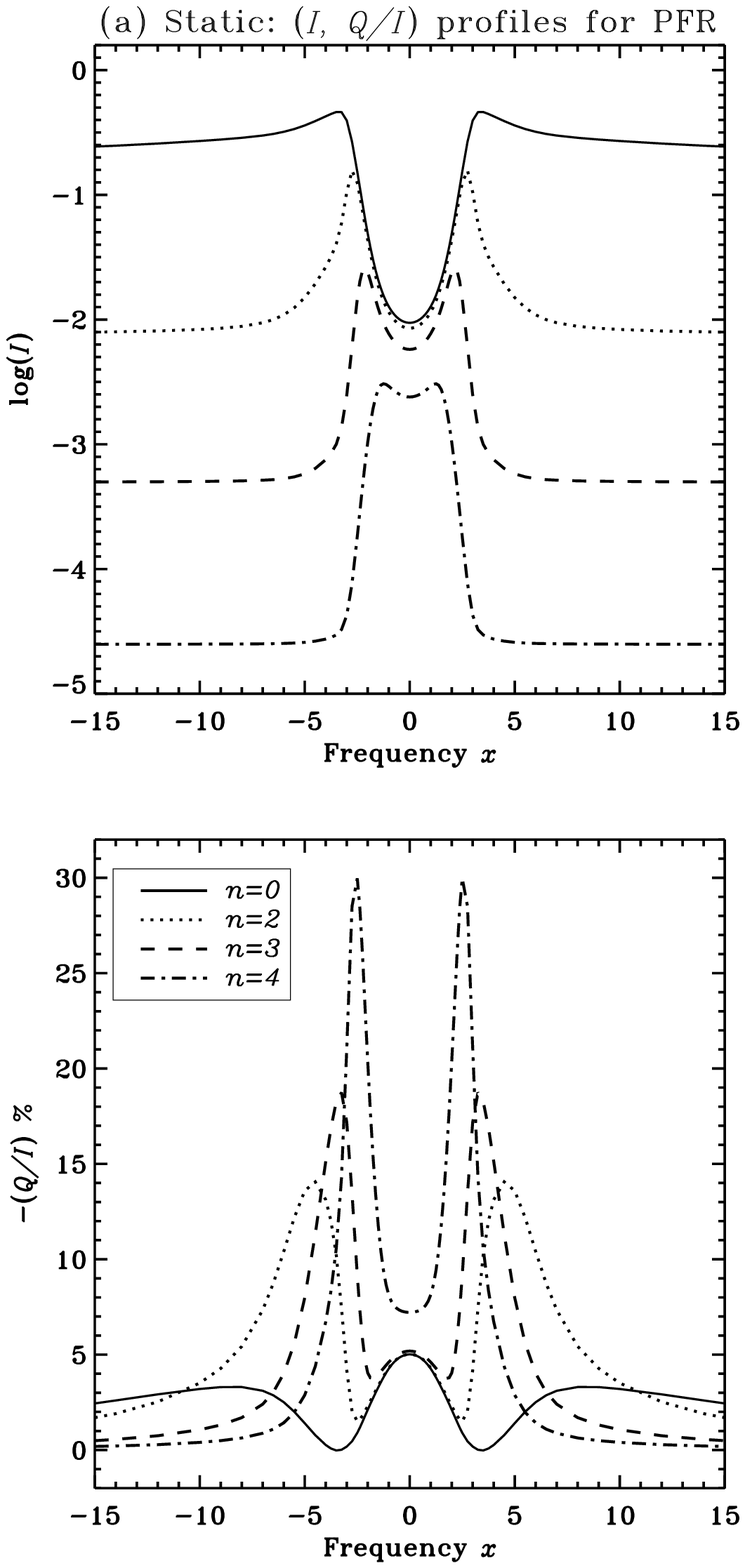}
\includegraphics[width=5cm,height=8.5cm]
{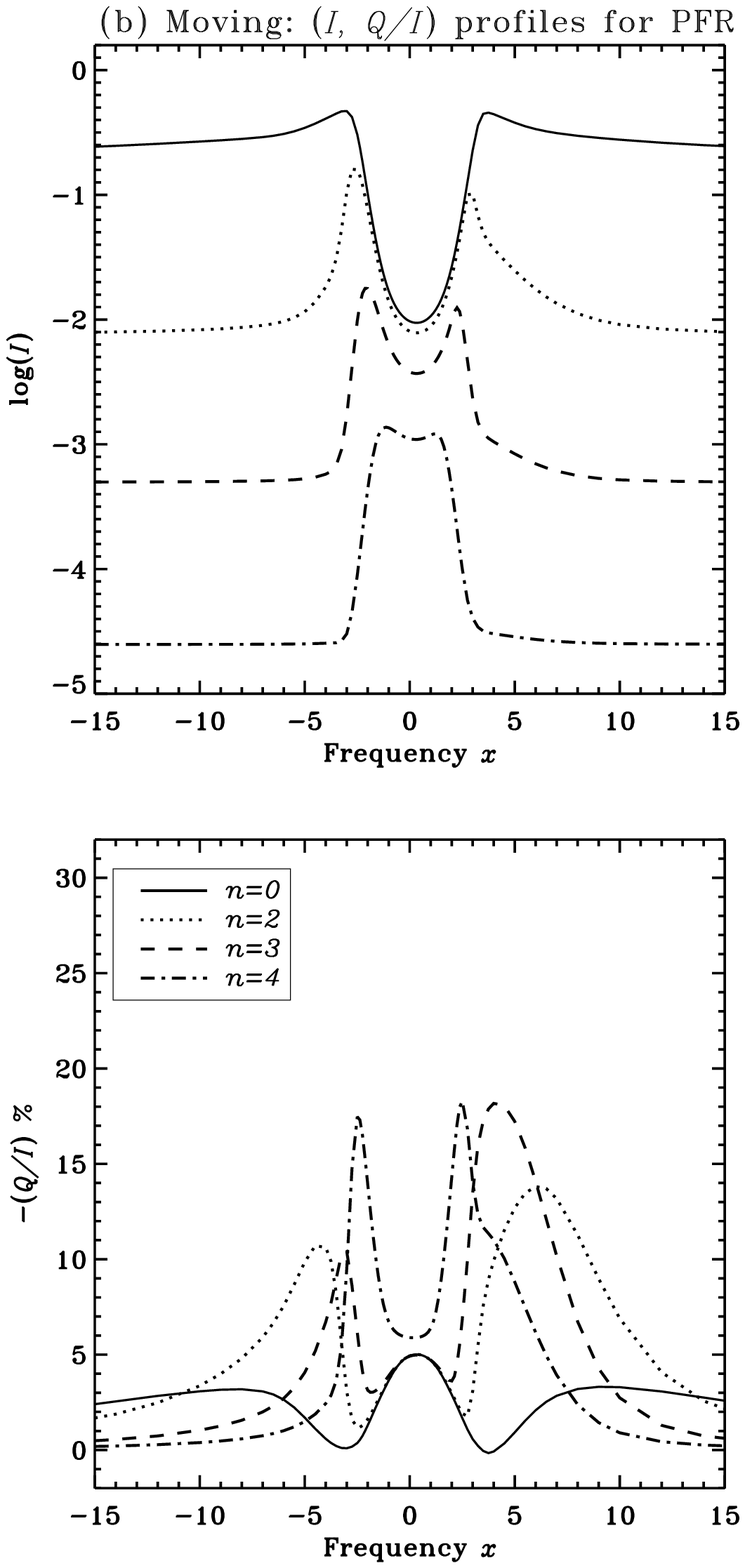}
 \caption{Emergent PFR $(I, Q/I)$ profiles from a spherically symmetric
static (panel a) and a constantly moving atmosphere
with $V(r)$=3 (panel b) for varying values of power law opacity index
$n$ (which are given in the inset box).}
 \label{indxn}
	 \end{center}
 \end{figure*}
%%%%%%%%%%%%%%%%%%%%%%%%%%%%%%%%%%%%%%%%%%%%%%%%%%%%%%%%%%%%%%%%%%%%%%
\subsection{Dependence on power law opacity index $n$}
Figure \ref{indxn} shows the emergent PFR $(I, Q/I)$ profiles 
from a spherically symmetric static (panel a) and a constantly moving 
atmosphere with $V(r)$=3 (panel b) for varying values of 
power law opacity index $n$. The density 
distribution in a spherical atmosphere generally obeys a power law type
of opacity distribution namely $\chi_{l,c}(r) \propto r^{-n}$.
Here we vary $n$ in the range 0 to 4. 
For $n=0$, we have a homogeneous sphere. It is well-known that sphericity 
effects do not develop fully in a homogeneous sphere and are seen only for 
$n > 1$ (see \citealt[]{kh1974}). Consequently the emission
contributions from the 
extended lobes are uniformly large for $n=0$ giving rise to an absorption 
profile in intensity.
As $n$ increases the scattering contribution from the 
extended lobes increases significantly
giving rise to self-absorbed emission profile for $n=2$, 3 and an
emission profile with extremely shallow line core absorption for $n=4$. 
Since with increasing $n$, the total opacity rapidly decreases as we move
outward, the intensity also drops as $n$ increases. For $n=0$, 2, and 3 the
$Q/I$ profile exhibits a typical triple peak structure. As $n$ increases,
the wing PFR peaks become more narrower and shift towards the line core
region due to the decrease in optical thickness of extended lobes. 
Due to dominate contributions from photons that are scattered small
number of times, the amplitude of wing PFR peaks also increases with $n$. 
For $n \geq 3$
the line core peak disappears as the medium becomes optically thin also at the
stellar core.
%As the $n$ value varies from
%-3 to 4 transition from absorption to emission takes place in 
%intensity profile.
%The polarization in the line core is not affected for varying $n$ between 
%-3 to 0 but the wings show the transition from negative to positive 
%polarization. For $n=2$ triple peak structure is prominently formed. Also
%the polarization in the core shifts towards the center and disappears 
%for $n=3$ and above. The wing PFR peaks also becomes narrower and shifts
%towards the central region. 
In the presence of velocity field intensity exhibits an
asymmetric profile about the line center for $n=2,$ 3, and 4, 
while the intensity for $n=0$ exhibits only a blue shift. 
Even the polarization 
profiles show asymmetry for $n=2,$ 3, and 4. For $n=4$ the $Q/I$
exhibits a strong Doppler dimming throughout the profile
as the corresponding $I$ profile is nearly in emission.
%%%%%%%%%%%%%%%%%%%%%%%%%%%%%%%%%%%%%%%%%%%%%%%%%%%%%%%%%%%%%%%%%%%%%%
\section{Conclusions}
The comoving frame method to solve the problem of polarized
PFR line formation in spherically symmetric atmospheres with velocity fields
is developed in \cite[]{megha2019}. Here we use this method to study the
effects of different model parameters on the linearly polarized line 
profiles in spherically symmetric static and moving atmospheres.
We have varied the model parameters in a wide range, one at a time,
to study their effects
on the line profile. We show that the line profiles exhibit 
a strong dependence on
each of the parameters selected for our study and thus help in understanding
the polarized lines formed in extended and expanding atmospheres. 
We show that the velocity fields
modify both the amplitude and shape of the $Q/I$ profiles. Such a 
modification is significant when the corresponding intensity profiles
exhibit a self-absorbed emission profile. This is because in this case both
Doppler brightening and dimming are simultaneously at play (see 
\citealt[]{megha2019} for details). In those cases where the intensity 
profiles are in emission, the corresponding $Q/I$ profiles exhibit a strong
Doppler dimming apart from a Doppler blue shift, when non-zero velocity fields 
are present in the line forming regions. It is important to note that strong
asymmetry between the blue and red wing PFR peaks do not exist if the 
intensity profiles are in emission or in pure absorption.
%%%%%%%%%%%%%%%%%%%%%%%%%%%%%%%%%%%%%%%%%%%%%%%%%%%%%%%%%
% \acknowledgments
\begin{acknowledgements}
 Megha would like to thank CSIR for the travel grant
TG/10856/19-HRD, Indian Institute of Astrophysics and organizers 
of SPW9 for the 
partial financial support to participate in the workshop.
KNN is thankful to the Director, Indian Institute of Astrophysics
for extending the research facilities. 
KNN also acknowledges the partial support by the organizers of 
SPW9 and by Dr. M. Bianda, Director, IRSOL that enabled him to 
participate in the workshop.
We acknowledge the use of the high performance computing 
facility at Indian Institute of Astrophysics.
\end{acknowledgements}

%
%\bibliographystyle{aa}
%\bibliography{lastname}
%%%%%%%%%%%%%%%%%%%%%%%%%%%%%%%%%%%%%%%%%%%%%%%%%%%%%%%%%%%%%%%%%%%%%%

\end{document}